\documentclass[pra,twocolumn,a4paper,superscriptaddress]{revtex4-2}

\usepackage{amsmath}
\usepackage{amssymb}
\usepackage{amsfonts}
\usepackage[english]{babel}
\usepackage{diagbox}
\usepackage{graphicx}
\usepackage{hyperref}
\usepackage{cleveref}
\usepackage{verbatim}
\usepackage{soul}
\usepackage{subcaption}
\usepackage{tabularx}
\usepackage{tikz}

\newcommand{\pd}[2]{\frac{\partial {#1}}{\partial{#2}}}

\newcommand{\abs}[1]{\left| {#1} \right|}
\newcommand{\modsq}[1]{\left| {#1} \right|^2}
\newcommand{\expect}[1]{\langle {#1} \rangle}
\newcommand{\ket}[1]{| {#1} \rangle}
\newcommand{\bra}[1]{\langle {#1} |}
\newcommand{\bracket}[1]{\left[ {#1} \right]}
\newcommand{\paren}[1]{\left( {#1} \right) }

\begin{document}
\title{Reference frame-independent model of a collective excitation atom interferometer}

\author{B.J.~{Mommers}}
\affiliation{Centre for Engineered Quantum Systems, The University of Queensland, St Lucia, Australia}
\email[]{b.mommers@uq.edu.au}
\affiliation{School of Mathematics and Physics, The University of Queensland, St Lucia, Australia}

\author{M.W.J.~{Bromley}}
\affiliation{School of Sciences, University of Southern Queensland, Toowoomba, Australia}
\affiliation{School of Mathematics and Physics, The University of Queensland, St Lucia, Australia}

\begin{abstract}
We theoretically analyze the operating principles of a proposed matter-wave Sagnac interferometer utilizing Bose-Einstein condensate (BEC) phonon modes as an interference medium. 
Previous work found that the orbital angular momentum phonon modes of a ring-trapped BEC are split in frequency by rotations, leading to a measurable rotation signal.
We develop an alternate description in which an imbalance in the counter-propagating modes' amplitudes (populations) is induced by the rotation of the system during condensation. 
This description gives analytic forms for the interferometic phase shift in 1D and is readily generalized to include mean-field interactions.
To validate our findings, we simulate a ring-trapped BEC Sagnac interferometer in one dimension and demonstrate that measurement of an unknown rotation rate can be performed using a modified analysis. 
Our simulation data show strong agreement with our analytic results, and we further employ simulations to explore and clarify the role of superfluidity in this matter-wave Sagnac interferometer.
\end{abstract}

\maketitle

\section{Introduction}
The Sagnac effect links the phase shift between waves counter-propagating within an enclosed loop with the external rotation of such a system~\cite{sagnac13a}.
This has been exploited to allow high-precision interferometric measurement of rotations, useful in inertial sensing and navigation
(for examples see Ref.~\cite{anderson94a}).
Current state-of-the-art Sagnac interferometers utilize counter-propagating light, with a large enclosed area to boost sensitivity~\cite{culshaw05a}.
Matter-wave systems appear to have a sensitivity advantage when considering the energy difference between optical and atomic systems~\cite{barrett14a}.
Despite this, matter-wave Sagnac interferometers are yet to surpass their optical counterparts in terms of precision.

Recent proposals have made use of advances in Bose-Einstein condensate (BEC) research and experimental techniques, providing an alternative medium to atomic beam-based and guided matter-wave interferometry schemes~\cite{wang05a,arnold06a,horikoshi07a,jo07a,burke09a,mcdonald13a,helm15a,haine18a,helm18a,zhou20a,moan20a,qin19a,gersemann20a,deppner21a,masi21a,moukouri21a,zhao22a,krzyzanowska22a,tomilin22a}.
In particular, proposals to measure the Sagnac phase shift of an interference pattern produced by counter-propagating orbital angular momentum (OAM) modes in a trapped BEC offer a way to utilize the high levels of control and coherence available in modern experimental systems~\cite{thanvanthri12a,marti15a,tomilin22a}.
This can be achieved through imparting optical OAM onto the condensate~\cite{thanvanthri12a} or by exciting standing wave collective excitations~\cite{marti15a} -- the latter being the focus of this work.

\begin{figure*}[tb]
\includegraphics[width=\linewidth]{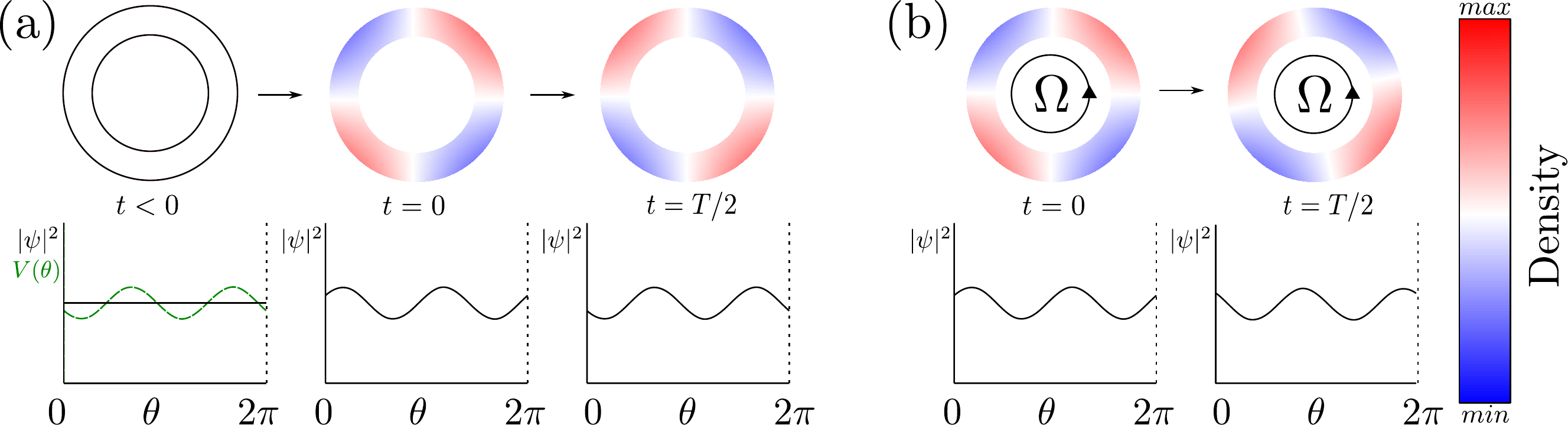}
\caption{Marti et al. proposal schematic illustration, with 2D top-down view (top) and 1D representation (bottom). (a) Begin with atoms confined to a ring with constant density and impose an optical potential with azimuthally-dependent modulations $V(\theta)$ while condensing to form a BEC with density modulations. At $t=0$, remove the optical potential and allow the density modulations to undergo standing wave oscillation, where at each half-cycle, the density modulation pattern is inverted. (b) In the presence of rotation, the modulation peaks shift position with each half-cycle.}
\label{fig:marti-schematic}
\end{figure*}

The use of collective excitation modes is expected to ameliorate technical difficulties often associated with BEC interferometric protocols that require condensate splitting or spin-dependent transitions~\cite{gustavson97a,torii00a,horikoshi07a,halkyard10a,petrovic13a,helm15a,helm18a,moan20a}.
A protocol for collective excitation Sagnac interferometry in a ring-trapped BEC has been proposed and tested for rotation sensing~\cite{marti15a}.
This protocol imprints a standing wave excitation on ring-trapped atoms through a weak azimuthally-modulated optical potential whilst cooling through the BEC transition.
By imposing this potential during condensate formation then releasing the BEC to freely evolve in the ring, rotation of the standing wave pattern is observable in the rotating laboratory frame, as illustrated in Fig.~\ref{fig:marti-schematic}.

Ref.~\cite{marti15a} analyze the standing wave rotation in terms of a frequency splitting between the standing wave's constituent counter-propagating travelling wave components in a three-mode model.
However, the frequency splitting effect requires observation in an inertial frame, something that cannot be achieved when attempting to experimentally measure unknown rotations.

This work is an alternate analysis of the theory in Ref.~\cite{marti15a}, in which we develop a frame-independent theory of operation for this protocol.
By analyzing the dynamics in the context of mode amplitudes we reveal multiple features of the interferometric design that were not previously apparent.
The splitting of counter-propagating modes in a rotating waveguide via the Sagnac effect has been exploited in optical systems to produce non-reciprocal propagation of light, analysed in terms of both mode amplitude- \cite{maayani2018a} and frequency-splitting \cite{huang18a,  jiao20a}.
For the purposes of matter-wave Sagnac interferometry for precision rotation sensing, we show that a mode amplitude-splitting analysis provides further insight into achieving high precision.

In Section \ref{sec:mode-imbalance} we describe the three-mode model and use it to derive the relationship between the rotation rate and standing-wave mode amplitudes, showing that rotation induces a splitting in the amplitudes of counter-propagating modes.
Section \ref{sec:simulations} details our numerical simulations of the interferometer, which show that rotation measurements can be performed by examining the evolution of the Fourier phase of the imprinted density modes.
In Section \ref{sec:numerical-ground-state} we show numerically that the imbalanced state is the ground state of the ring system in the presence of both rotation and an imprinting potential.
In Section \ref{sec:mean-field-excitations} we extend our analysis to low-temperature BECs described by the Gross-Pitaevskii equation (GPE), and examine practical considerations of our results.

\section{\label{sec:mode-imbalance} Three-mode model}
In this section we introduce the three-mode model and use it to derive the Sagnac phase shift for the ring geometry in both inertial and rotating frames.
In the rotating frame we directly relate the mode imbalance to the rotation rate.

Our initial analysis considers a simple model of the proposed phonon-mode interferometer: a non-interacting system confined to a one-dimensional ring. This ideal phonon-mode interferometer with wavefunction $\psi(\theta)$ populates only three angular momentum eigenstates,
\begin{align}
\psi(\theta) &= \phi _0 + \phi _{+l} + \phi _{-l} ~, \label{eq:3mode-psi}\\
\phi _0 &= a_0 ~,\label{eq:3mode-phi0}\\
\phi _{l} &= \left(a + \Delta\right)\ e^{il\theta} ~, \label{eq:3mode-phi+}\\
\phi _{-l} &= \left(a-\Delta\right)\ e^{-il\theta} \label{eq:3mode-phi-}~.
\end{align}
Here $\phi_k$ is an angular momentum eigenstate with quantized circulation $k \in \{0,l,-l\}$, $a$ is the mean amplitude of the of the $k=\pm l$ modes, $\Delta$ is the mode splitting, and $\theta$ is the positional parameter (angle) around the 1D ring. The wavefunction is normalized such that $\int \modsq{\psi} dr= 1$. 
Therefore $\phi _0$ is the constant background on which the angular momentum modes (i.e. phonon modes) with OAM quantum number $\pm l$ are imprinted.
Non-zero values of $\Delta$ result in an imbalance between counter-propagating mode amplitudes.

\subsection{\label{sec:subsection-inertial-frame} In the inertial frame}

To begin, we first note that Ref.~\cite{marti15a} assumes an equal superposition of counter-propagating modes, i.e. $\Delta = 0$. 
Our analysis proceeds without this assumption. 
For each mode (\cref{eq:3mode-phi0,eq:3mode-phi+,eq:3mode-phi-}), the time evolution operator $U_t = e^{-i\hat{H}t/\hbar}$ is given by,
\begin{align}
U_t \phi_0 &= a_0 \ e^{-iE_0 t/\hbar} ~, \label{eq:Ut_phi0} \\
U_t \phi_{l} &= (a + \Delta) \ e^{i(l\theta - E t/\hbar)} ~, \label{eq:Ut_phi+} \\
U_t \phi_{-l} &= (a - \Delta) \ e^{-i(l\theta + E t/\hbar)} \label{eq:Ut_phi-} ~.
\end{align}

Where $E_0$ is the energy of the background mode, and $E = E_l = l^2 \hbar^2/2mr^2$ is the degenerate energy of the counter-propagating modes, with $m$ representing particle mass and $r$ the ring's radius, which we both set to 1 without loss of generality.

The time-dependent density profile can be decomposed into the sum of a standing wave and a travelling wave oscillating on a time-independent background,
\begin{align}
\modsq{\psi(t)} =~&a_0 ^2 + (a + \Delta)^2 + (a - \Delta)^2 \nonumber \\
                &+ 2 (a + \Delta) (a - \Delta) \cos  (2l\theta) \nonumber \\
 				&+ 4 a_{0} \Delta \cos [l \theta - (E-E_0)t/\hbar] \nonumber \\
 				&+ 2 a_{0} (a - \Delta) \left( \cos [l \theta - (E-E_0)t/\hbar] \nonumber \right.\\
                &+ \left. \cos [l \theta + (E-E_0)t/\hbar] \right) \label{eq:swdecomp}
\end{align}
Note when $\Delta = 0$ the travelling wave component of Eqn.~ \ref{eq:swdecomp}, $4 a_{0} \Delta \cos [l \theta - (E-E_0)t/\hbar]$, is zero, consistent with a non-rotating standing wave state.
The rotation rate is determined through the evolution of the complex phase shift in the single-frequency Fourier transform of the density profile,
\begin{align}
\varphi(t) &= \arctan \bracket{\int \modsq{\psi(t)} e^{-il\theta} d\theta}~,
\label{eq:fourier-decomp}
\end{align}
for a single mode $l$.

From Eqns.~\ref{eq:swdecomp}~and~\ref{eq:fourier-decomp}, the phase shift in the inertial frame takes the form,
\begin{align}
\varphi(t) &= \arctan \bracket{- \frac{\Delta}{a} \tan((E-E_0)t/\hbar)} \label{eq:inertial-phase}~.
\end{align}
In the limit $\frac{\Delta}{a} \rightarrow 1$, Eqn.~\ref{eq:inertial-phase} gives a constant phase gradient - a linear phase accumulation consistent with a pure travelling wave.
Conversely, taking the limit $\Delta \rightarrow 0$ results in a constant zero phase gradient - as expected for a pure standing wave.

Considering Eqn.~\ref{eq:inertial-phase} at stroboscopic measurement times,
\begin{equation}
 \tau_s = \frac{n\pi\hbar}{E-E_0} ,
 \label{eq:strobe}
\end{equation}

where $n$ is a natural number encoding the number of oscillations since $t=0$, the phase shift is always zero.
The rotation of the density profile relative to the measurement frame (in this case the inertial frame) is given by the time derivative of the phase shift at stroboscopic times -- when the standing wave component is at its maximum amplitude,
\begin{align}
\frac{d\varphi}{dt} &= -\frac{\Delta (E-E_0)}{\hbar a}
\label{eq:inertial-omega} ~.
\end{align}
In the inertial frame this depends only on the mode imbalance and choice of excited mode number.
In a sensing application, the measurement frame is expected to rotate with the system at an unknown rate.
By analysing the phase shift $\varphi(t)$ in the rotating frame, we can determine the relationship between rotation rate $\Omega$ and mode imbalance $\Delta$.

\begin{figure*}[tb]
\centering
\includegraphics[width=\linewidth]{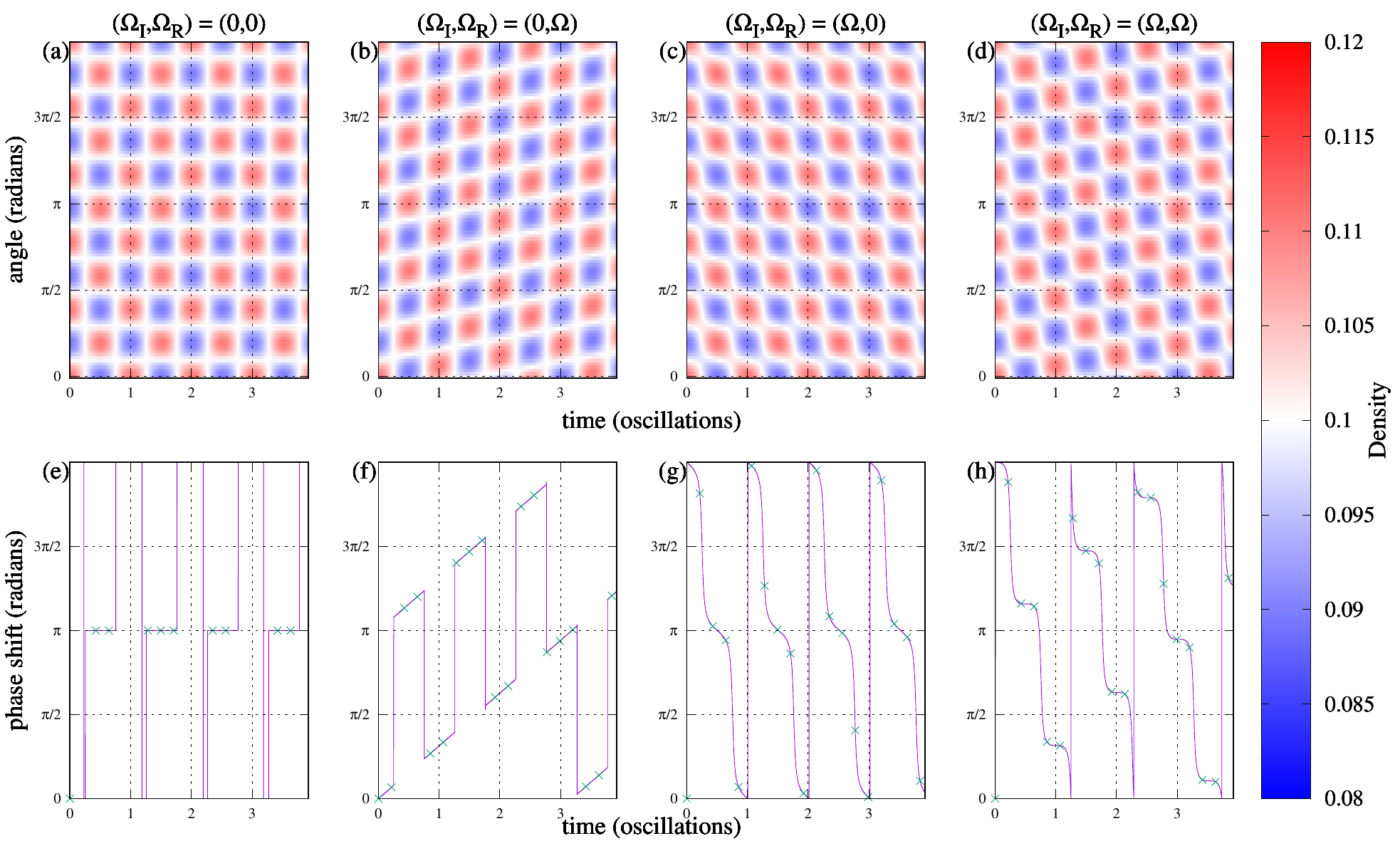}
\caption{Space-time density profiles and extracted phase for different rotating reference frame configurations from simulations without interactions.
    (a)--(d): space-time density profiles for the ideal 1D ring system.
    Each case is labelled with the rotation rate of the system during imaginary time evolution ($\Omega_I$) which corresponds to condensate formation, and the real-time rotation rate ($\Omega_R$) that corresponds to the free evolution of the system once the imprinting potential is removed. 
    Dotted vertical lines indicate full oscillations of the standing wave component from $t=0$, whilst dotted horizontal lines indicate angular positions around the ring, to guide the eye. 
    (e)--(h): plots of the extracted phase for each of the simulation conditions matching the density profile plot immediately above.
    The purple curve is from simulations using the ground state found via imaginary time, while the green points are from simulations that numerically optimize Eqn.~\ref{eq:Hexpect}. 
    The numerics in Ref.~\cite{marti15a} correspond to subfigures (b) and (f), whilst a physical experiment necessarily corresponds to (d) and (h).
}
\label{fig:spacetime-phasext-g0}
\end{figure*}

\subsection{\label{sec:subsection-rotating-frame} In the rotating frame}
In the rotating frame, we can perform a similar analysis as above with the inclusion of a time-dependent coordinate transform,
\begin{align}
\theta  &\rightarrow \theta' + \Omega t ~.
\end{align}
Repeating the calculation of the time-dependent density, the Fourier transform result for a rotating system is given by,
\begin{align}
\int \modsq{\psi(\theta',t)} e^{-il\theta'} d\theta' =~ &a_0 \left( a + \Delta \right) e^{-i (E-E_0)t/\hbar~+il \Omega t } \nonumber \\
    &+ a_0 \left( a - \Delta \right) e^{i (E-E_0)t/\hbar~+il \Omega t }  ~.
\end{align}
The phase angle subsequently depends on the rotation rate,
\begin{widetext}
\begin{align}
\varphi(t) = \text{arctan}~&\left[ \frac{-\Delta \sin[(E-E_0) t/\hbar] \cos(l \Omega t)  + a \cos[(E-E_0)t/\hbar] \sin(l \Omega t)}{\Delta \sin[(E-E_0) t/\hbar] \sin(l \Omega t)  + a \cos[(E-E_0)t/\hbar] \cos(l \Omega t)} \right] ~.
\label{eq:rotating-phase}
\end{align}
\end{widetext}
For $\Omega = 0$, this reduces to Eqn.~\ref{eq:inertial-phase}.
At stroboscopic times $\tau_s$ (Eqn.~\ref{eq:strobe}) we find explicit dependence of the phase shift on the rotation rate,
\begin{align}
\varphi(\tau_s) &= \frac{n\, \pi\, \hbar\, l\, \Omega}{E-E_0} ~.
\label{eq:rotating-stroboscopic-phase-shift}
\end{align}
This phase shift increases with each oscillation due to the $n$-dependence of Eqn.~\ref{eq:rotating-stroboscopic-phase-shift}. At exactly $t=\tau_s$, the rotation of the density profile is zero relative to the rotating frame. At other times, it follows the general form of the phase shift time derivative in the rotating frame,
\begin{align}
\frac{d\varphi}{dt} &= l \Omega - \frac{\Delta (E-E_0)}{\hbar a}  ~.
\label{eq:rotating-phase-gradient}
\end{align}
As $\frac{d\phi}{dt} = 0$ at stroboscopic measurement times, we can directly relate the mode imbalance $\Delta$ to the rotation rate $\Omega$ using Eqn.~\ref{eq:rotating-phase-gradient},
\begin{align}
\frac{\Delta}{a} (E-E_0) = l \hbar \Omega ~. \label{eq:rotation-rate}
\end{align}
This result demonstrates that a rotating system develops an imbalance in the amplitudes of imprinted counter-propagating OAM modes which is intrinsically linked to the rate of rotation -- with non-zero rotation there is always a non-zero imbalance.

Experimentally, the rotation rate is measured by imaging the atomic density at one or more stroboscopic measurement times and determining the Fourier phase shift.
The linear gradient between stroboscopic phase shifts according to Eqn.~\ref{eq:rotating-stroboscopic-phase-shift} is proportional to $l\Omega$ -- therefore an unknown rotation rate can be measured in a frame where the laboratory is also rotating.

\section{\label{sec:simulations} Numerical Simulations}
To test and extend our analytic results, we simulate the dynamics of a one-dimensional ring-trapped condensate in two parameter regimes: the phonon regime and the mean-field regime. In the phonon regime, excitations are of sufficiently small amplitude that inter-particle interactions are negligible -- allowing simulation of dynamics using the Schr\"odinger equation for wavefunction $\psi(\theta,t)$,
\begin{align}
-i\hbar \pd{\psi (\theta,t)}{t}  &= \paren{\frac{-\hbar ^2}{2mr^2} \pd{^2}{\theta ^2} + V(\theta,t)} \psi(\theta,t) ~. \label{eq:schrodinger}
\end{align}
In the mean-field regime, interactions are non-negligible and described by the Gross-Pitaevskii equation (GPE),
\begin{align}
-i\hbar \pd{\psi (\theta,t)}{t}  &= \paren{\frac{-\hbar ^2}{2mr^2} \pd{^2}{\theta ^2} + V(\theta,t) + g\modsq{\psi(\theta,t)}} \psi(\theta,t) ~, \label{eq:gpe}
\end{align}
where $g$ is the interaction strength parameter, $r$ is the radius of the ring, $V(\theta,t)$ the time-dependent imprinting potential, and in the GPE $\psi(\theta,t)$ is the 1D order parameter for the bosonic field.
Simulation is performed using a three-point Crank-Nicolson method on a grid of 105 points.
We simulate the full protocol: condensation (via imaginary time evolution) in an $|l|=5$ imprinting potential, and removal of the imprinting potential at $t=0$ for free evolution of the system.
The Fourier components at select frequencies corresponding to OAM modes $l = [-15,15]$ are calculated at each timestep, including the phase shift extracted from the density profile.
Rotation of the system was independently set for the imaginary time and real time portions of the simulation via the rotation parameters $\Omega_I, \Omega_R$, respectively.
Density plots and extracted phase shifts for the non-interacting system are presented in Fig.~\ref{fig:spacetime-phasext-g0}.

Several features of our numerical results support our analytic solution from Sec.~\ref{sec:mode-imbalance}.
In the inertial frame $(\Omega_I = \Omega_R = 0)$ the phase shift at stroboscopic times is zero, as shown in Fig.~\ref{fig:spacetime-phasext-g0}~(e).
Similarly, Fig.~\ref{fig:spacetime-phasext-g0} shows the time derivative of the phase shift in the rotating frame $(\Omega_I = \Omega_R = \Omega)$ is zero, and the phase shift accumulates by a fixed amount proportional to $l\Omega$ each oscillation, as predicted by Eqn.~\ref{eq:rotating-stroboscopic-phase-shift}. 

Figure~\ref{fig:spacetime-phasext-g0} highlights the differences in both phase shift and density oscillations under different rotation conditions.
The rotating-frame analysis in Sec.\ref{sec:mode-imbalance} corresponding to Fig.~\ref{fig:spacetime-phasext-g0}~(d) and (h) matches the conditions of an experimental measurement of an unknown rotation -- the condensate is prepared under rotation and freely evolves in the ring trap under rotation.

\section{\label{sec:numerical-ground-state} Numerical calculation of ground state}
The protocol in Ref.~\cite{marti15a} requires an imprinting potential during condensation to form the initial state.
According to the three-mode model, the ground state in the imprinting potential will depend on model parameters $a_0,a,\Delta$.
In this section, we calculate the expectation value of the imprinting Hamiltonian using a three-mode state and compare its numerically-calculated minimum value to the ground state found using imaginary time evolution.

\begin{figure*}[tb]
\centering
\begin{subfigure}{0.45 \linewidth}
\includegraphics[width= \linewidth]{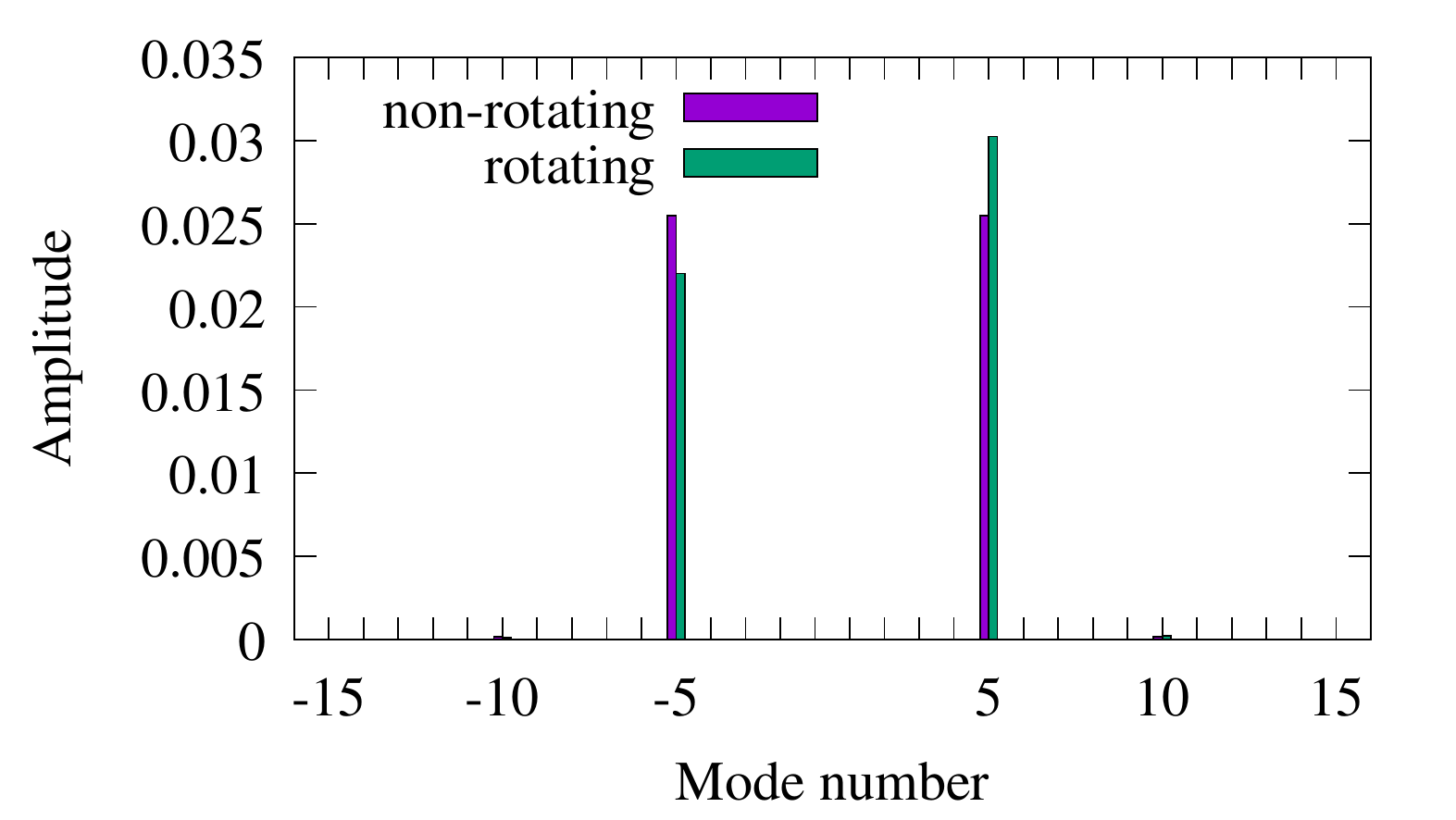}
\caption{$g=0$}
\label{fig:t0-g0-modes}
\end{subfigure}
\begin{subfigure}{0.45 \linewidth}
\includegraphics[width= \linewidth]{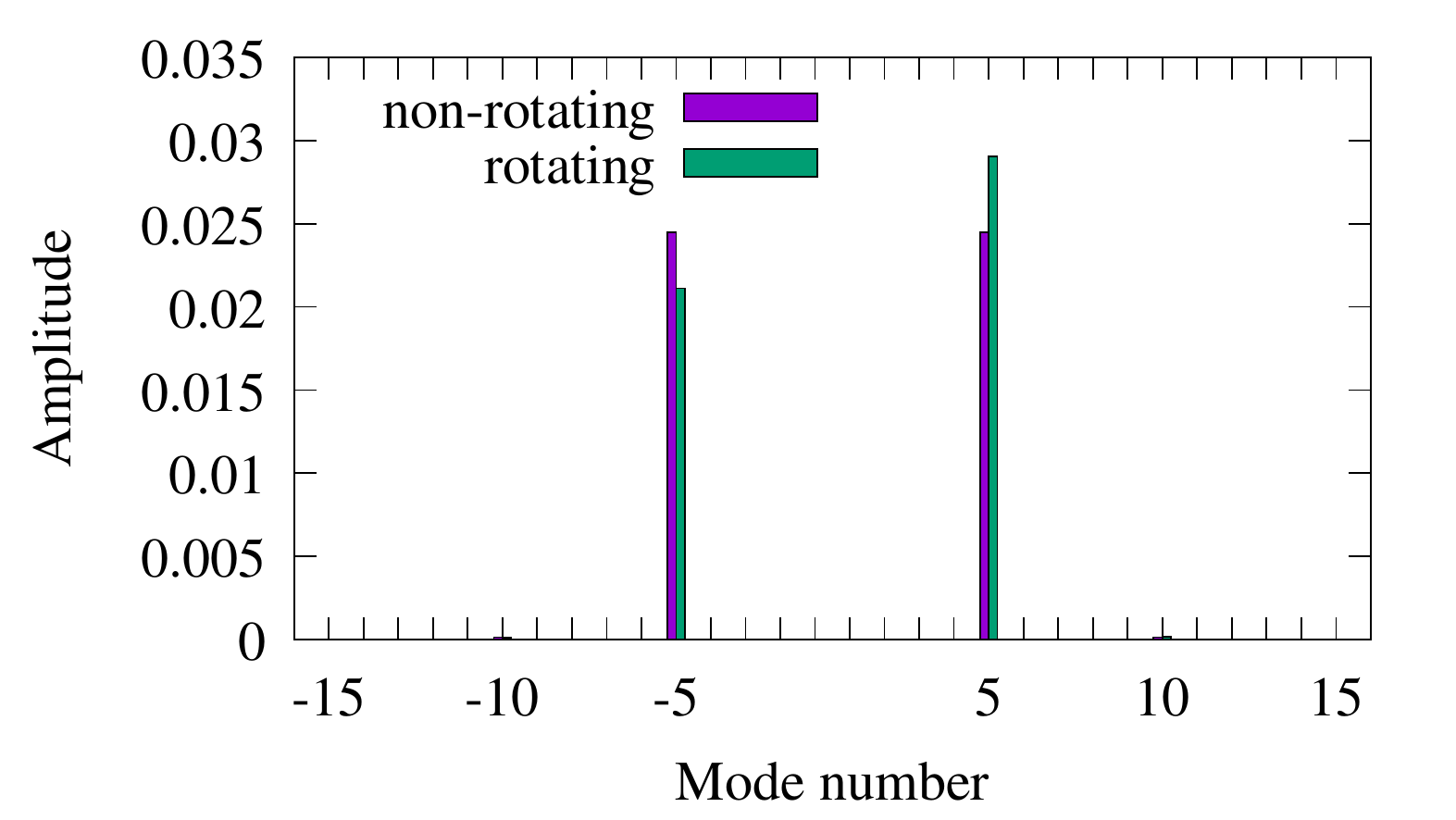}
\caption{$g=1$}
\label{fig:t0-g1-modes}
\end{subfigure}
\begin{subfigure}{0.45 \linewidth}
\includegraphics[width= \linewidth]{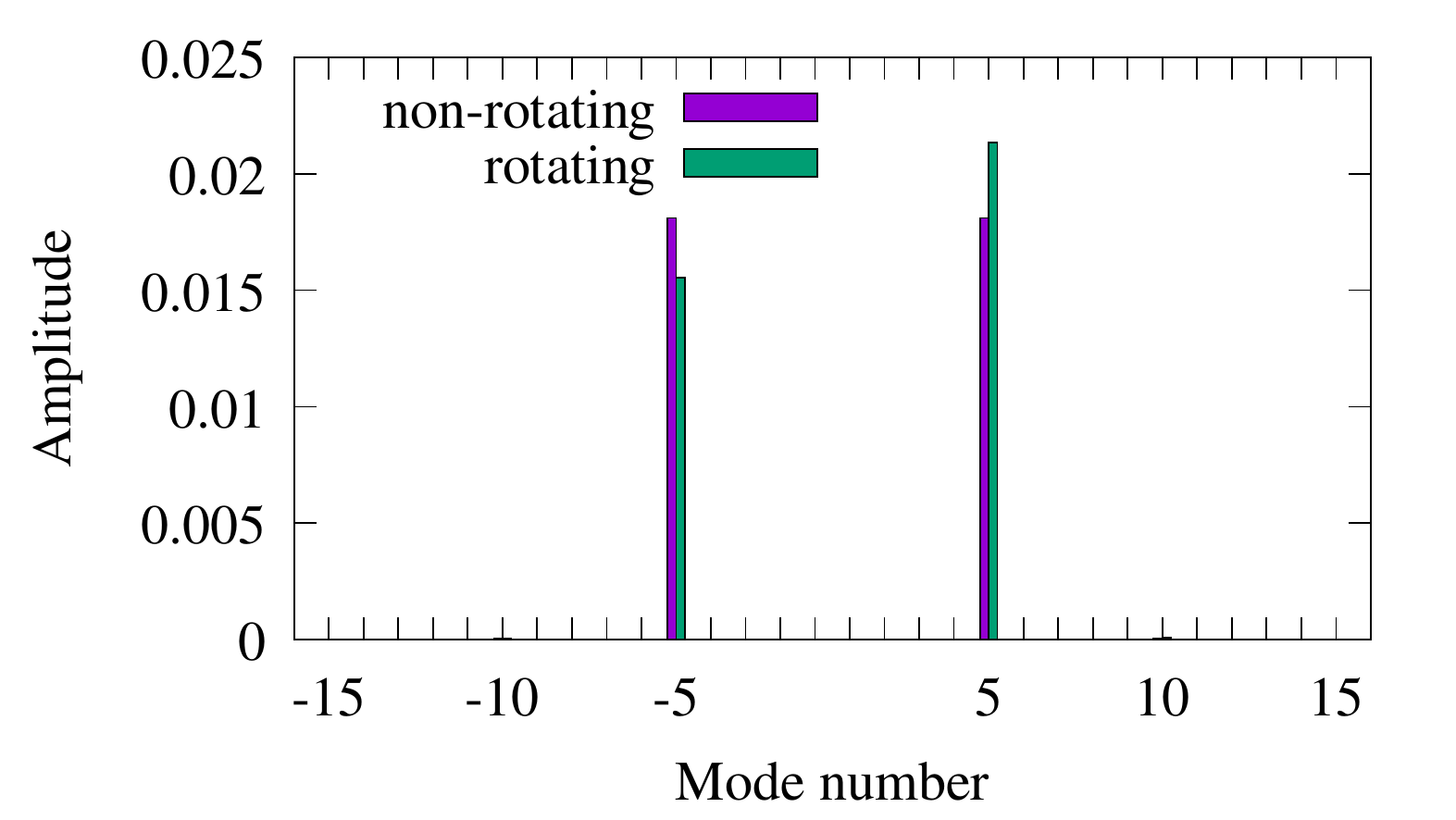}
\caption{$g=10$}
\label{fig:t0-g10-modes}
\end{subfigure}
\begin{subfigure}{0.45 \linewidth}
\includegraphics[width= \linewidth]{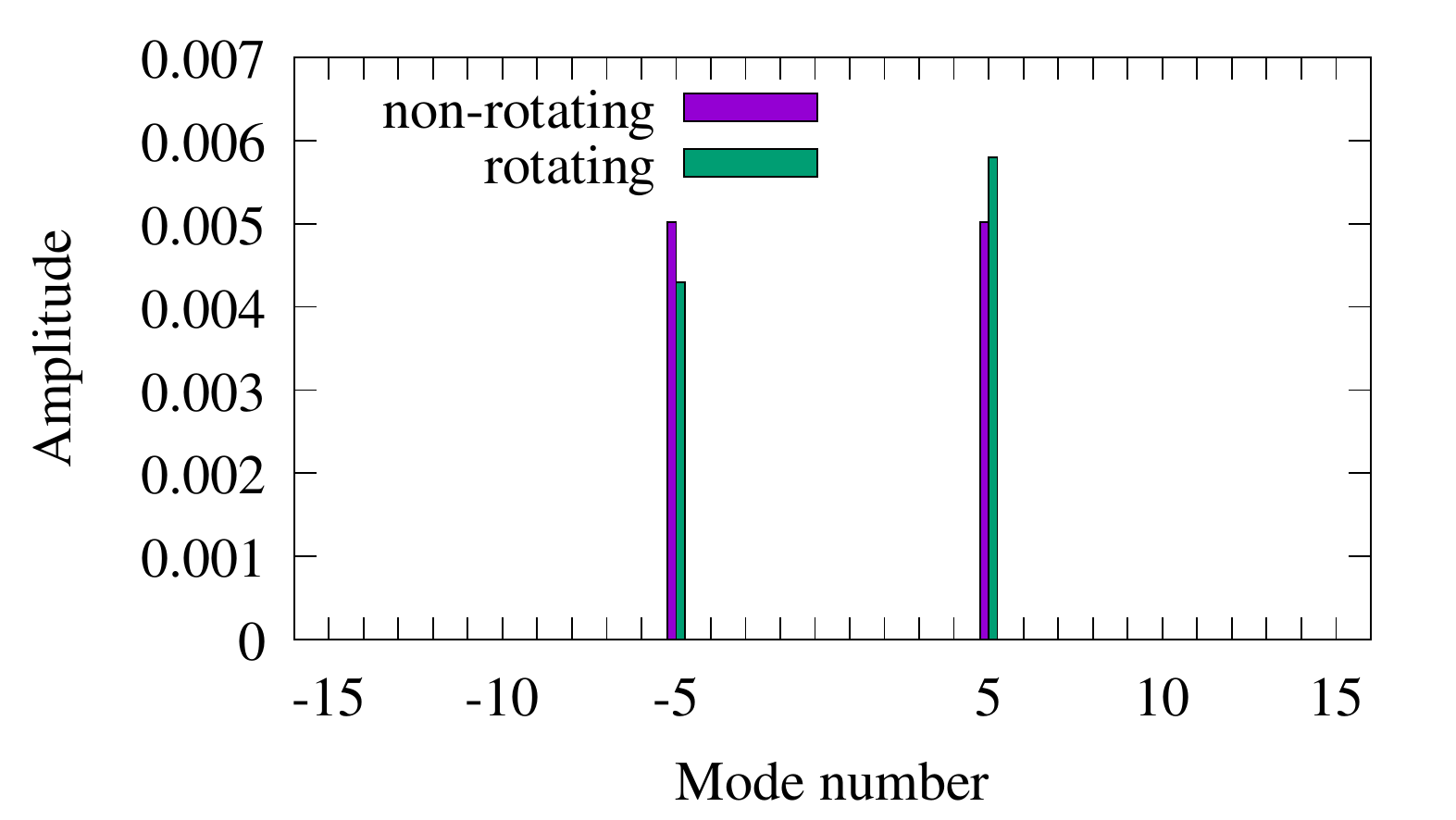}
\caption{$g=100$}
\label{fig:t0-g100-modes}
\end{subfigure}
\caption{Mode decomposition of the simulated state following imaginary time evolution with and without rotation for increasing interaction strength $g$. The mode splitting is clear between the non-rotating and rotating cases. The dominant $l=0$ mode has been omitted for visual clarity. Note that the amplitude in the non-rotating case is not the mean amplitude for the corresponding rotating case due to coupling between parameters $a$ and $\Delta$ in Eqns.~\ref{eq:Hexpect} and \ref{eq:Hexpect-g}. As $g$ is increased, we see a suppression of the imprinted mode.}
\label{fig:mode-decomp-g}
\end{figure*}

We simulate the 1D particle-in-a-ring Hamiltonian with imprinting potential in a rotating frame,
\begin{align}
\hat{H} = -\frac{\hbar^2}{2 m R^2} \pd{^2}{\theta^2} + [V_0 -\alpha \cos(l\theta)] + i \hbar \Omega \pd{}{\theta} ~.
\end{align}
Here $m$ is the particle mass, $R$ the ring radius, $\Omega$ the rotation rate, and $V=(1-\alpha \cos(l\theta))$ the imprinting potential with modulation amplitude $\alpha$.
We choose as our \emph{ansatz} $\psi$ from Eqn.~\ref{eq:3mode-psi}, with normalization factor $\mathcal{N} = [2 \pi (a_0 ^2 + 2 a^2 + 2\Delta ^2)]^{-1/2}$.
We then calculate the expectation value of the Hamiltonian, $\expect{\hat{H}}$,
\begin{align}
    \expect{\hat{H}} &= \frac{\frac{\hbar ^2 l^2}{2mR^2} \left( 2 a^2 + 2 \Delta ^2 \right) + 4 l \Omega a \Delta - 2 \alpha a_0 a}{a_0^2 + 2 a^2 + 2 \Delta ^2} ~. \label{eq:Hexpect}
\end{align}
Full details of the calculation are available in Appendix~\ref{app:numerics}.
The ground state of the system is defined by the values of $a_0,a,\Delta$ that minimize this expectation value for given values of the ring radius, rotation rate, and imprinting potential amplitude.
Note that for non-zero rotation $\abs{\Omega} >0$, there is an explicit coupling of $a$ and $\Delta$, therefore the ground state of a non-rotating system will have a different value for $a$ than that of an identical system that is rotating. 
This can be seen by careful examination of Fig.~\ref{fig:mode-decomp-g}.
As the normalization factor couples the derivatives with respect to our three main parameters, we use a numerical gradient-descent method to obtain values for $a_0,a,\Delta$ and compare to the converged ground state obtained by simulating the system in 1D with imaginary time evolution.
The main result of this comparison is shown in Fig.~\ref{fig:spacetime-phasext-g0}(e--h), where the extracted phase of both the three-mode model and imaginary time simulations are shown with strong agreement.

\section{\label{sec:mean-field-excitations} Mean-field excitations}

To investigate the effects of imprinting deeper modulations, we consider the 1D Gross-Pitaevskii equation (GPE) from Eqn.~\ref{eq:gpe} to describe mean-field condensate dynamics.

As we increase the interaction parameter $g>0$, the amplitude of the primary ($l$) mode Fourier component decreases as the repulsive interactions of the mean-field potential suppress modulations, as shown in Fig.~\ref{fig:mode-decomp-g}.

\begin{figure*}[tb]
\centering
\includegraphics[width=\linewidth]{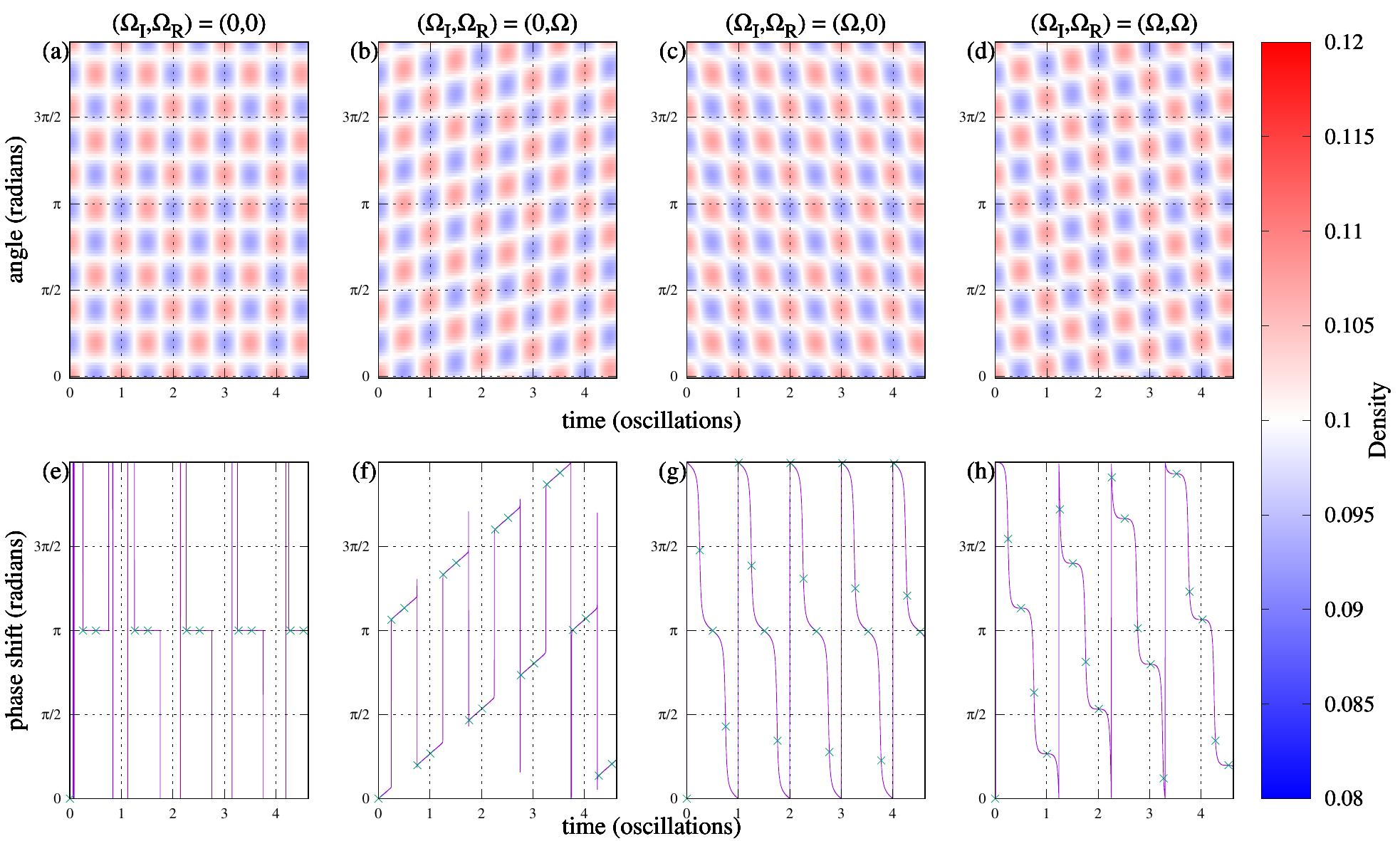}
\caption{Same as Fig.~\ref{fig:spacetime-phasext-g0} but with mean field interaction strength $g=10$. Note the lighter color in the density plots (a)-(d) due to interactions suppressing the imprinted mode amplitudes. The temporal oscillation frequency is also increased due to the extra energy in the system.
The Fourier phase evolution of the three-mode model (green crosses, subfigures (e)-(h)) agrees with that of the ground state found through imaginary time (purple curve).}
\label{fig:spacetime-phasext-g10}
\end{figure*}

The ground state for the interacting case is determined by using the GPE Hamiltonian from Eqn.~\ref{eq:gpe}.
This gives a modified expectation value to that in Sec.~\ref{sec:numerical-ground-state}, though it can be numerically minimized in the same way as Eqn.~\ref{eq:Hexpect},
\begin{align}
  \expect{\hat{H}} = &~\frac{1}{a_0^2 + 2 a^2 + 2 \Delta ^2} \bigg(  \frac{\hbar ^2 l^2}{2mR^2} \left( 2 a^2 + 2 \Delta ^2 \right) + 4 l \Omega a \Delta \nonumber \\
  &- 2 \alpha a_0 a + g [ 8a_0^2 a^2 + (a^2 - \Delta^2)^2 a^4 ]  \bigg) ~.
\label{eq:Hexpect-g}
\end{align}
As shown in Fig.~\ref{fig:spacetime-phasext-g10}, the interacting ground state result provides strong agreement between optimized parameters and those found via imaginary time evolution in numerical simulations.

\begin{figure}[tb]
\centering
\includegraphics[width=\linewidth]{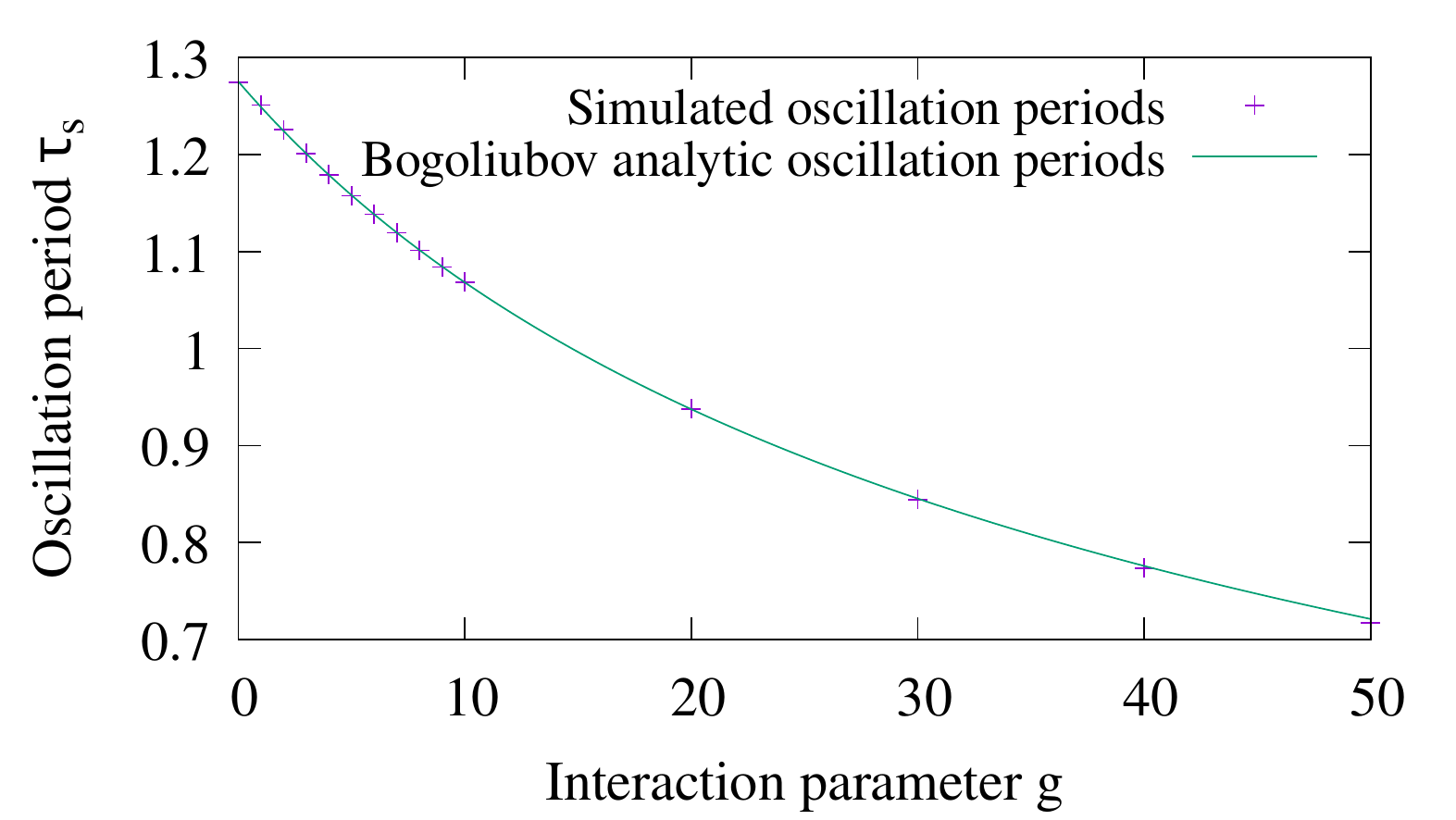}
\caption{Dependence of the temporal oscillation period on interaction strength $g$. Points are the oscillation periods extracted from numerical simulations, the curve corresponds to the analytic Bogoliubov mode energy from Eqn.~\ref{eq:bogoliubov-energy}. Natural units are set such that $\hbar = m = c = 1$.}
\label{fig:g-dependence}
\end{figure}

An additional feature of Fig.~\ref{fig:spacetime-phasext-g10} when viewed in comparison to Fig.~\ref{fig:spacetime-phasext-g0} is the dependence of the density oscillation period on the interaction strength.
In Fig.~\ref{fig:g-dependence} we compare the oscillation period of simulations with varying interaction strengths and find that in the mean-field approximation the oscillation frequency is dependent on the Bogoliubov mode energy,
\begin{align}
\epsilon(p) &= \sqrt{\paren{\frac{p^2}{2m}}^2 + \frac{p^2 g n }{m}} ~, \label{eq:bogoliubov-energy}
\end{align}
where $p$ is the mode's momentum and $n$ the atomic density.

\subsection{\label{sec:hess-fairbank-effect} The Hess-Fairbank effect}
\begin{figure}
\centering
\includegraphics[width=\linewidth]{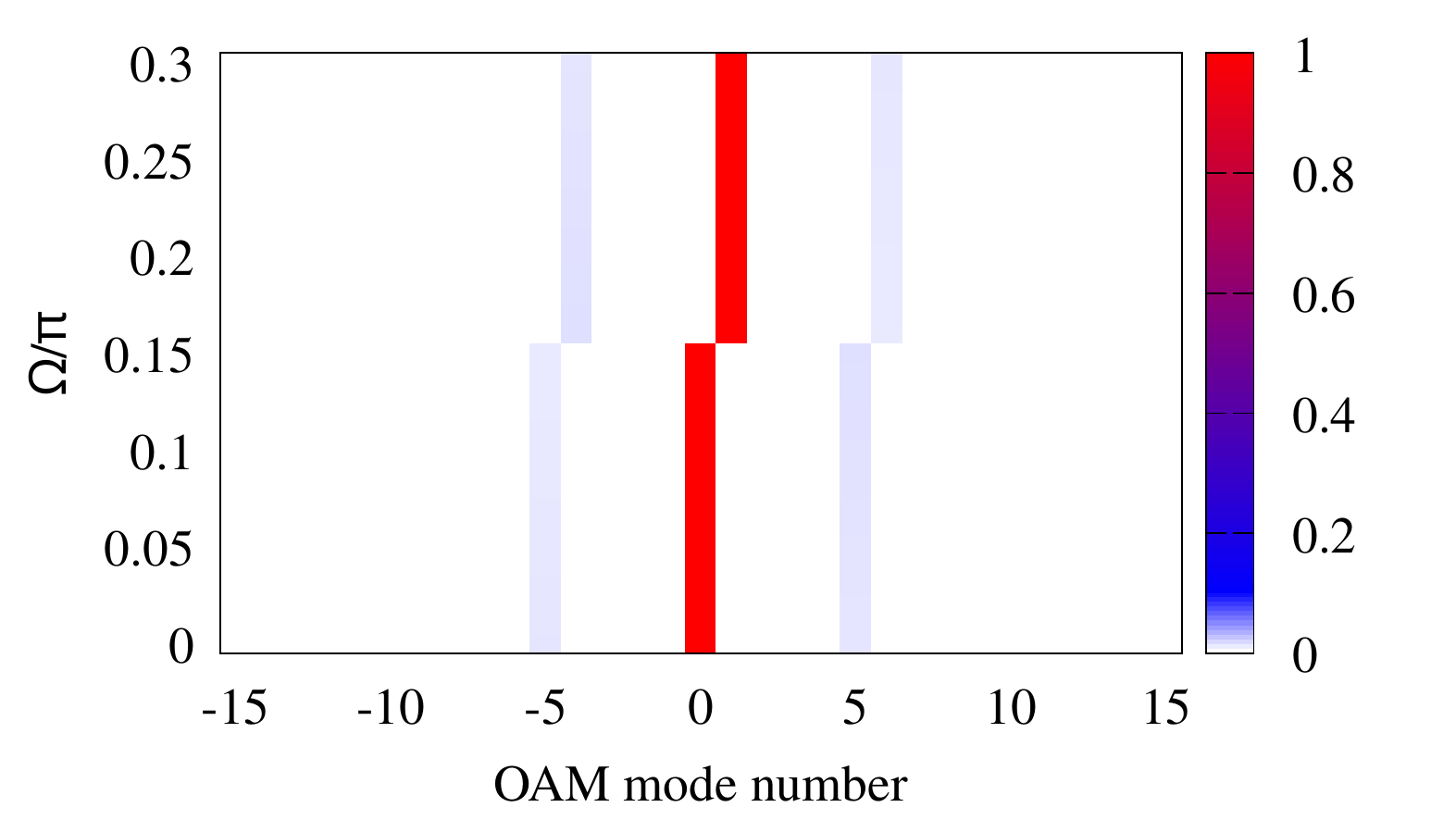}
\caption{Heatmap of normalized mode amplitudes for quantized angular momentum modes formed under rotation. Imaginary time evolution of an initial Gaussian state under external rotation at $\Omega$ produces the expected three-mode state at low rotation rates, and induces a global rotation of the system at higher rates.}
\label{fig:hess-fairbank}
\end{figure}

The Hess-Fairbank effect describes the formation of a non-rotating superfluid condensate in a sufficiently-slowly rotating container~\cite{hess67a}.
It is in this slowly-rotating regime that the three-mode model is obtained exactly, as shown in Fig.~\ref{fig:hess-fairbank} -- higher external rotation rates induce a global circulation in the condensed state, shifting the mode occupation.
This corresponds to a global phase factor in the wavefunction, which does not alter the mode structure of the density, nor the evolution of the Fourier phase (except for the expected scaling with rotation rate).
This is explored further in Appendix~\ref{app:hess-fairbank}.

A condensate formed in a rotating Sagnac interferometer is not irrotational as described by the Hess-Fairbank effect.
In the slowly-rotating regime, the rotation of the imprinted density profile indicates flow without global circulation.
This is where classical fluid analogies break down, as they cannot describe the superposition state of the condensate.
A simple experimental test of the Hess-Fairbank effect in the collective excitation interferometer can be performed using a rotating imprinting potential to simulate different external rotation rates as in Fig.~\ref{fig:hess-fairbank} without requiring rotation of the entire experimental apparatus.

\subsection{\label{sec:experimental-considerations} Experimental considerations}
The rotating reference frame and mean-field effects described above impact the feasibility of the original interferometric scheme.
In a typical inertial sensing application, the time-dependence of the phase shift requires stroboscopic measurements to be precisely timed for maximum accuracy.
This is complicated by the dependence of the stroboscopic measurement time on the interaction strength -- which in an experiment is related to the atom number, and cannot be precisely controlled between iterations.
Using the Bogoliubov mode energy as a guide to ideal measurement time can mitigate this where interactions are sufficiently weak.

\section{Conclusion}
In conclusion, the three-mode model with imbalanced mode amplitudes is a frame-independent model for studying the rotation signal in a phonon-mode Sagnac interferometer.
We have shown that this model is accurate in describing the one-dimensional behavior of this system, and that absolute rotation measurement is possible.
The three-mode state is not irrotational per the Hess-Fairbank effect even at low rotation rates, and we have suggested a simple experiment to demonstrate this.

Given the results of this work, the ideal rotation measurement scheme involves measuring the extracted phase at as close to the first stroboscopic measurement time as possible.
We have shown that there are multiple subtleties in reference frame and interaction effects that must be considered when designing ring-trapped interferometric schemes.

\begin{acknowledgments}
The authors would like to thank
M.C.~Kandes, R.~Carretero-Gonzalez, S.A.~Haine, T.W.~Neely, and M.J.~Davis for useful discussions,
and especially thank H.~Rubinsztein-Dunlop for her support.
This research was funded by the
(Australian) Defence Science and Technology Next Generation Technologies Fund (QT95).
This research was partially funded through an Australian
Research Council Future Fellowship (ARC, FT100100905),
supported by the Australian Research Council Centre of Excellence for
Engineered Quantum Systems (EQUS, CE170100009),
and Australian Government Research Training Program Scholarships.
\end{acknowledgments}

\appendix
\section{Numerical ground state calculation} \label{app:numerics}
In order to numerically determine the three-mode model parameters that define the ground state in a rotating ring system, we calculate the expectation value of the Hamiltonian for numerical minimization.
The 1D particle-in-a-ring Hamiltonian with imprinting potential is given in a rotating frame by:
\begin{align}
\hat{H} = -\frac{\hbar^2}{2 m R^2} \pd{^2}{\theta^2} + [V_0 -\alpha \cos(n\theta)] + i \hbar \Omega \pd{}{\theta}
\end{align}
Where $m$ is the particle mass, $R$ the ring radius, $\Omega$ the rotation rate, and $V=(1-\alpha \cos(n\theta))$ the imprinting potential with modulation amplitude $\alpha$.
We choose as our \emph{ansatz} $\psi$ from the three-mode model, with normalization factor $(2 \pi (a_0 ^2 + 2 a^2 + 2\Delta ^2))^{-1/2}$.
We can then determine the expectation value of the Hamiltonian, $\expect{\hat{H}}$:
\begin{align}
\expect{\hat{H}} = &\int _0 ^{2\pi} \psi^* \hat{H} \psi \ d\theta \nonumber \\
                 = &\frac{1}{2 \pi \paren{\modsq{a_0} + 2 \modsq{a} + 2 \modsq{\Delta}}} \nonumber \\
                 &\times \int _0 ^{2 \pi} \bracket{V_0 -\alpha \cos(m \theta)} \modsq{a_0} \nonumber \\
                 &+ \paren{\frac{\hbar ^2 l^2}{2 m R^2} + \bracket{V_0 - \alpha \cos(n \theta)} - \hbar l \Omega} \modsq{a+\Delta} \nonumber \\
                 &+ \paren{\frac{\hbar ^2 l^2}{2 m R^2} + \bracket{V_0 - \alpha \cos(n \theta)} + \hbar l \Omega} \modsq{a-\Delta} \nonumber \\
                 &+ \bigg[ \paren{\frac{\hbar ^2 l^2}{2 m R^2} + \bracket{V_0 - \alpha \cos(n \theta)} - \hbar l \Omega}  \nonumber \\
                 &\qquad \times   a_0 ^* \paren{a+\Delta} e^{il \theta} \bigg] \nonumber \\
                 &+ \bigg[ \paren{\frac{\hbar ^2 l^2}{2 m R^2} + \bracket{V_0 - \alpha \cos(n \theta)} + \hbar l \Omega} \nonumber \\
                 &\qquad \times a_0 ^* \paren{a-\Delta} e^{-il \theta} \bigg] \nonumber \\
                 &+ \paren{V_0 - \alpha \cos(n \theta)} \paren{a+\Delta}^* a_0\ e^{-il \theta}\nonumber \\
                 &+ \paren{V_0 - \alpha \cos(n \theta)} \paren{a-\Delta}^* a_0\ e^{il \theta}\nonumber \\
                 &+ \bigg[ \paren{\frac{\hbar ^2 l^2}{2 m R^2} + \bracket{V_0 - \alpha \cos(n \theta)} - \hbar l \Omega} \nonumber \\
                 & \qquad \paren{a-\Delta}^* \paren{a+\Delta} e^{-2il \theta} \bigg] \nonumber \\
                 &+ \bigg[ \paren{\frac{\hbar ^2 l^2}{2 m R^2} + \bracket{V_0 - \alpha \cos(n \theta)} + \hbar l \Omega} \nonumber \\
                 &\qquad \paren{a+\Delta}^* \paren{a-\Delta} e^{-2il \theta} \bigg] \ d\theta   \label{eq:Hexpint}
\end{align}
This integral is simplified by noting that all terms with a single oscillatory factor integrate to zero, leaving only the non-oscillatory terms, and terms with a product of $\cos(n \theta)$ and an exponential.
The latter can be calculated as follows:
\par
To determine the contribution of the imprinting potential modulations to the integral in Equation \ref{eq:Hexpint}, we calculate the potential matrix element for generic OAM eigenstates $\ket{k} = \psi _k = (2\pi)^{-1/2} e^{i k \theta}$:
\begin{align}
    \bra{l} V_{m} \ket{n} =~&\int _0 ^{2\pi} \psi ^* _l (\theta)\ [V_0 - \alpha \cos(m\theta)]\ \psi _n (\theta)\  d\theta \nonumber \\
                          =~&\int _0 ^{2\pi} \psi ^* _l (\theta)\ V_0 \ \psi _n (\theta)\  d\theta \ \nonumber \\
                          &- \frac{\alpha}{2\pi} \int _0 ^{2\pi} \psi ^* _l (\theta)\ \cos(m\theta)\ \psi _n (\theta)\  d\theta \nonumber \\
                          =~&V_0 \ \delta _{l,n} - \frac{\alpha}{2\pi} \int _0 ^{2\pi}  \cos(m\theta)\ e^{-i(l-n)\theta} \  d\theta \nonumber \\
                          =~&V_0 \ \delta _{l,n} - \frac{\alpha}{2\pi} \int _0 ^{2\pi}  \frac{1}{2} \big( \cos[(l-n+m)\theta] \nonumber \\
                          & \hspace{2.5cm} + \cos[(l-n-m)\theta] \big) \nonumber \\
                          & \hspace{2.5cm} - \frac{i}{2} \big( \sin[(l-n+m)\theta] \nonumber \\
                          & \hspace{2.5cm} - \sin[(l-n-m)\theta] \big) \  d\theta      
\end{align}
The integral is non-zero if and only if either $l-n+m=0$, or $l-n-m=0$. In either of these cases, the matrix element reduces to:
\begin{align}
    \bra{l} V_{m} \ket{n} &= - \frac{\alpha}{2}
\end{align}
For the three-mode model, there are four such terms, resulting in the $-2\alpha a_0 a$ term in the final expression for the expectation value of the Hamiltonian:
\begin{align}
    \expect{\hat{H}} =~&\frac{1}{a_0^2 + 2 a^2 + 2 \Delta ^2} \nonumber \\
    &\times \bigg( \frac{\hbar ^2 l^2}{2mR^2} \paren{2 a^2 + 2 \Delta ^2} + 4 \hbar l \Omega a \Delta - 2 \alpha a_0 a \bigg)   \label{eq:app_Hexpect}
\end{align}
This process is repeated using the Gross-Pitaevskii equation to obtain the nonlinear expectation value presented in the main text.

\section{The Hess-Fairbank effect} \label{app:hess-fairbank}
First observed in liquid Helium \cite{hess67a}, the Hess-Fairbank effect describes the formation of a non-rotating superfluid in a container rotating sufficiently slowly during the transition to the superfluid phase.
In a simply-connected cylindrical container of radius $R$, rotation proportional to $\frac{n\hbar}{mR^2}$ is sufficient to generate $n$ vortices of quantized circulation.
This places an upper limit on the rotation rate of such a container such that the ground state has zero circulation.
\par
In the ring geometry of the phonon-mode interferometer, it is straightforward to calculate the rotation rate at which an energy crossing between two OAM modes occurs for a non-interacting superfluid by equating their eigenenergies:
\begin{align}
    E_l &= E_k \\
    \frac{\hbar^2 l^2}{2mR^2} - \hbar l \Omega &= \frac{\hbar^2 k^2}{2mR^2} - \hbar k \Omega \\
    \frac{\hbar}{2mR^2} (l^2 - k^2) &= \Omega (l-k) \\
    \Omega &= \frac{\hbar}{mR^2} \frac{l^2 - k^2}{2 (l-k)} \label{eq:crossing_rotation_rate}
\end{align}
For the $l^{th}$ mode and the mode directly adjacent (in the direction of rotation), the crossing occurs at $\Omega = \frac{\hbar}{mR} ~ \frac{2l+1}{2}$.
However, in the three-mode model each mode does not shift to the adjacent mode, instead there is a patterned shift where the $\pm l^{\text{th}}$ mode shifts to the $\mp (l + 1)^{\text{th}}$ mode for rotation about the positive $z$-axis (or shifts to the $\mp (l - 1)^{\text{th}}$ mode for rotation about the negative $z$-axis), as shown in Table \ref{tab:mode-shift}.

\onecolumngrid

\begin{table}[h!]
\renewcommand{\arraystretch}{1.2}
\setlength{\extrarowheight}{20pt}
\begin{tabularx}{\linewidth}{X|X X X X X X X X X X X X}
\diagbox[width=1.4cm,height=1.8cm]{$~k$}{$l~$} & ~$-5$   & $-4$   & $-3$   & $-2$   & $-1$   & $0$    & $1$    & $2$    & $3$    & $4$    & $5$   & $6$    \\ 
\hline
$-5$                 &      & ~$-\frac{9}{2}$ & $-4$   & $-\frac{7}{2}$ & $-3$   & $-\frac{5}{2}$ & $-2$   & $-\frac{3}{2}$ & $-1$   & $-\frac{1}{2}$ & $0$   &  \hl{$\frac{1}{2}$}~ \\
$-4$                 & ~$-\frac{9}{2}$ &      & $-\frac{7}{2}$ & $-3$   & $-\frac{5}{2}$ & $-2$   & $-\frac{3}{2}$ & $-1$   & $-\frac{1}{2}$ & $0$    & $\frac{1}{2}$ & $1$    \\
$-3$                 & ~$-4$   & $-\frac{7}{2}$ &      & $-\frac{5}{2}$ & $-2$   & $-\frac{3}{2}$ & $-1$   & $-\frac{1}{2}$ & $0$    & $\frac{1}{2}$  & $1$   & $\frac{3}{2}$  \\
$-2$                 & ~$-\frac{7}{2}$ & $-3$   & $-\frac{5}{2}$ &      & $-\frac{3}{2}$ & $-1$   & $-\frac{1}{2}$ & $0$    & $\frac{1}{2}$  & $1$    & $\frac{3}{2}$ & $2$    \\
$-1$                 & ~$-3$   & $-\frac{5}{2}$ & $-2$   & $-\frac{3}{2}$ &      & $-\frac{1}{2}$ & $0$    & $\frac{1}{2}$  & $1$    & $\frac{3}{2}$  & $2$   & $\frac{5}{2}$  \\
$0$                  & ~$-\frac{5}{2}$ & $-2$   & $-\frac{3}{2}$ & $-1$   & $-\frac{1}{2}$ &      & \hl{$\frac{1}{2}$}~  & $1$    & $\frac{3}{2}$  & $2$    & $\frac{5}{2}$ & $3$    \\
$1$                  & ~$-2$   & $-\frac{3}{2}$ & $-1$   & $-\frac{1}{2}$ & $0$    & $\frac{1}{2}$  &      & $\frac{3}{2}$  & $2$    & $\frac{5}{2}$  & $3$   & $\frac{7}{2}$  \\
$2$                  & ~$-\frac{3}{2}$ & $-1$   & $-\frac{1}{2}$ & $0$    & $\frac{1}{2}$  & $1$    & $\frac{3}{2}$  &      & $\frac{5}{2}$  & $3$    & $\frac{7}{2}$ & $4$    \\
$3$                  & ~$-1$   & $-\frac{1}{2}$ & $0$    & $\frac{1}{2}$  & $1$    & $\frac{3}{2}$  & $2$    & $\frac{5}{2}$  &      & $\frac{7}{2}$  & $4$   & $\frac{9}{2}$  \\
$4$                  & ~$-\frac{1}{2}$ & $0$    & $\frac{1}{2}$  & $1$    & $\frac{3}{2}$  & $2$    & $\frac{5}{2}$  & $3$    & $\frac{7}{2}$  &      & $\frac{9}{2}$ & $5$    \\
$5$                  & ~$0$    & \hl{$\frac{1}{2}$}~  & $1$    & $\frac{3}{2}$  & $2$    & $\frac{5}{2}$  & $3$    & $\frac{7}{2}$  & $4$    & $\frac{9}{2}$  &     & $\frac{11}{2}$  \\
$6$                  & ~$\frac{1}{2}$  & $1$    & $\frac{3}{2}$  & $2$    & $\frac{5}{2}$  & $3$    & $\frac{7}{2}$  & $4$    & $\frac{9}{2}$  & $5$    & $\frac{11}{2}$ &   \\
\hspace{0.1mm}
\end{tabularx}
\caption{Rotation rates as multiples of $\frac{\hbar}{mR^2}$ that produce energy crossings from OAM mode $k$ to mode $l$; determined using Equation \ref{eq:crossing_rotation_rate}. Highlighted entries show the mode shifting for the three-mode model studied in the text. Note that the observed shift is not due to adjacent modes coupling for $l=\pm 5$.}
\label{tab:mode-shift}
\end{table}

\begin{figure}[t!]
    \centering
    \includegraphics[width=0.9\linewidth]{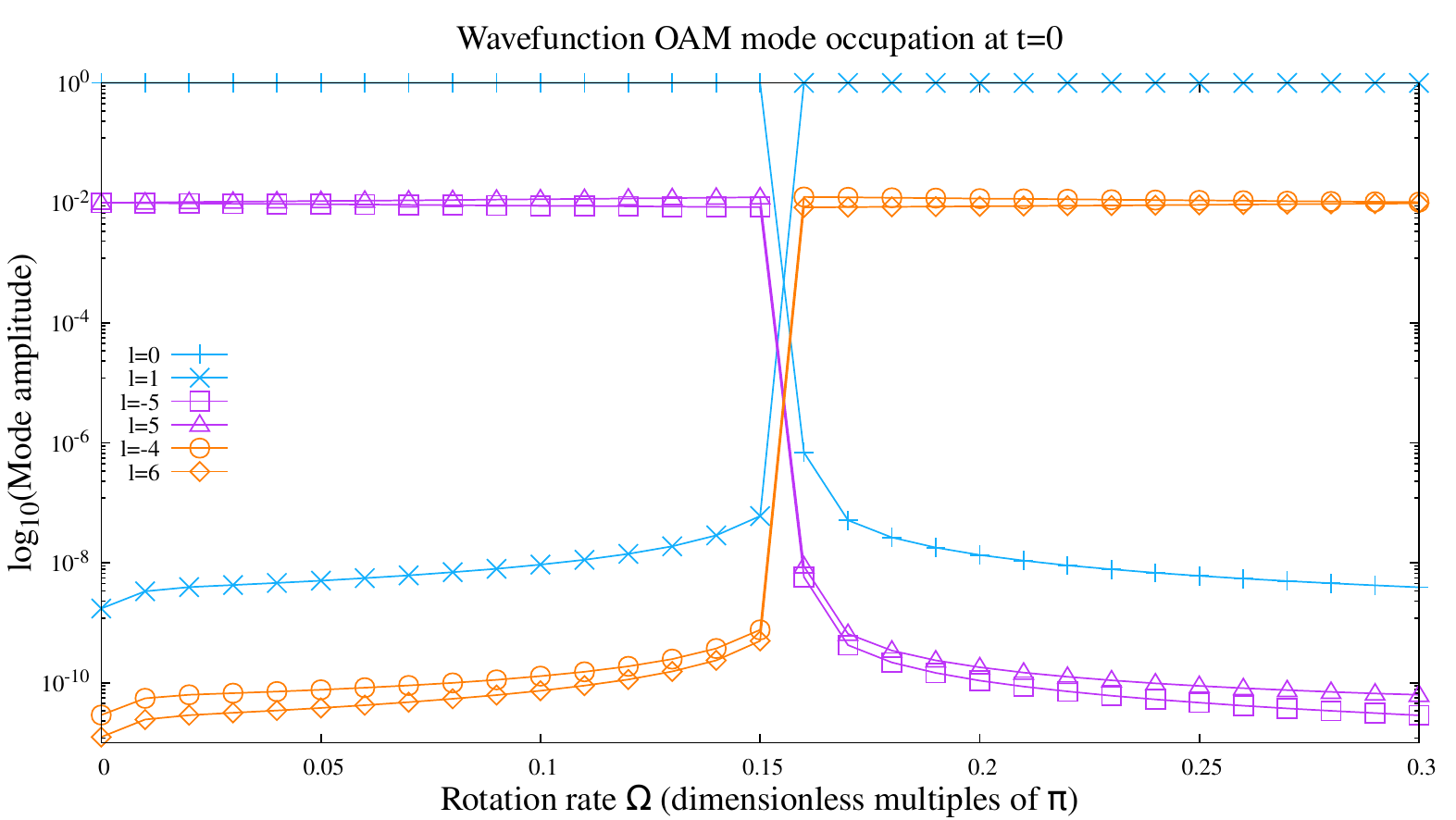}
    \caption{Occupation of select orbital angular momentum (OAM) modes after condensation into an $l=5$ imprinting potential for increasing external rotation rate during condensation. Values below $10^{-6}$ are equivalent to zero due to a finite convergence threshold.}
    \label{fig:hess-fairbank-1D-logscale}
\end{figure}

\begin{figure}[b!]
    \centering
\includegraphics[width=0.9\linewidth]{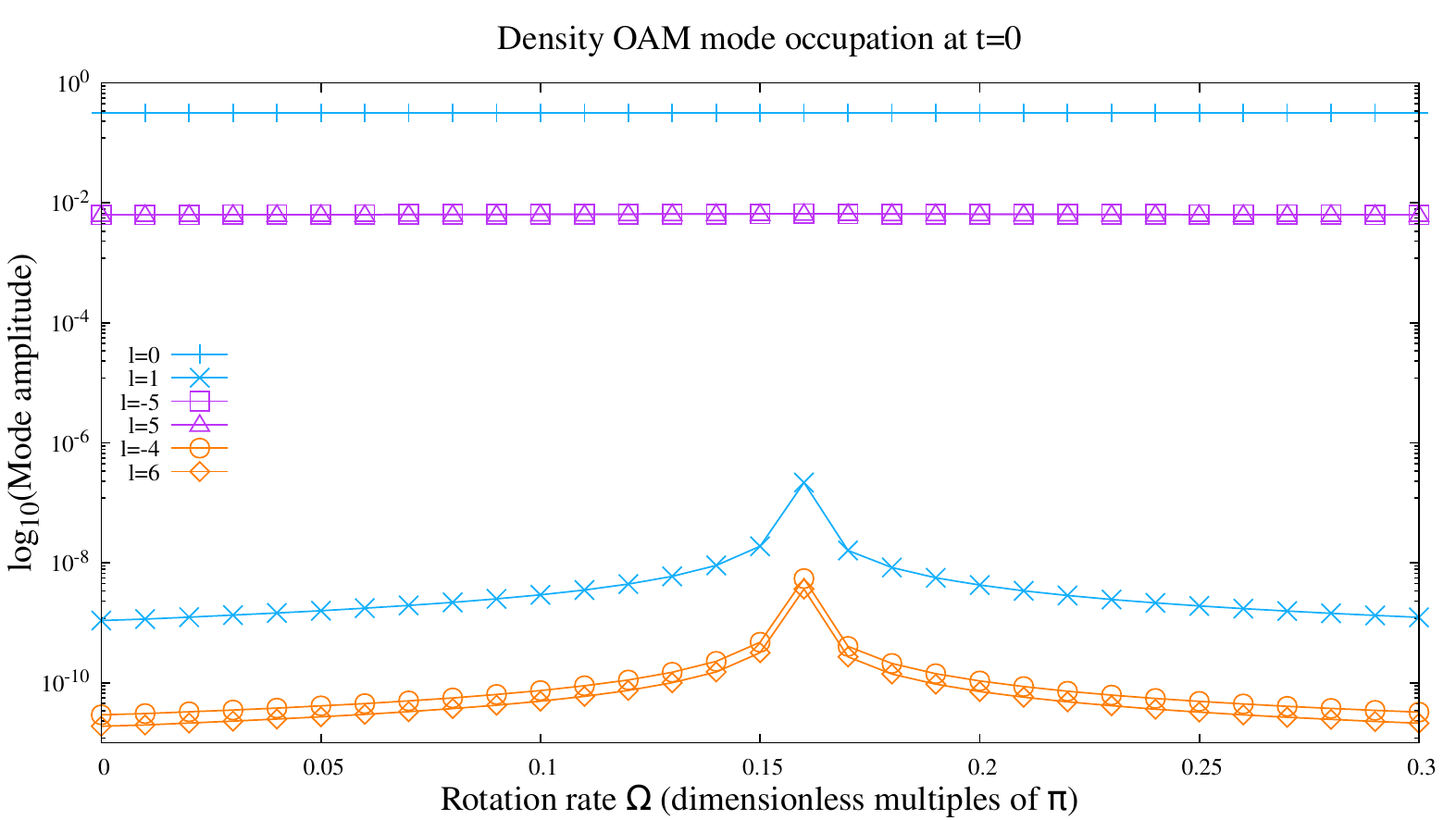}
    \caption{Amplitudes of select density Fourier modes after condensation into an $l=5$ imprinting potential for increasing external rotation rate during condensation. Values below $10^{-6}$ are equivalent to zero due to a finite convergence threshold.}
    \label{fig:hess-fairbank-1D-density-logscale}
\end{figure}

\twocolumngrid

\pagebreak

\end{document}